# MODELING THE CONSEQUENCES OF TONGUE SURGERY ON TONGUE MOBILITY

S. Buchaillard[1, 3], M. Brix[2], P. Perrier[1] & Y. Payan[3]

[1] ICP /GIPSA-lab, UMR CNRS 5216, INP Grenoble, F38031 Grenoble - France
[2] University Hospital, Grenoble - France
[3] TIMC-IMAG, UMR CNRS 5525, Université Joseph Fourier, Grenoble – F38700 La Tronche - France

## INTRODUCTION

Tongue surgery can have severe consequences on tongue mobility and tongue deformation capabilities. This can generate strong impairments of three basic functions of human life, namely mastication, swallowing and speech, which induce a noticeable decrease of the patients' quality of life [1,2].

In the line of works carried out in predictive medicine to set up systems of computer aided surgery, this paper presents the current achievements of a long term project aiming at predicting and assessing the impact of tongue and mouth floor surgery on tongue mobility. The ultimate objective of this project is the design of a software with which surgeons should be able (1) to design a 3D biomechanical model of the tongue and of the mouth floor that matches the anatomical characteristics of each patient specific oral cavity, (2) to simulate the anatomical changes induced by the surgery and the possible reconstruction, and (3) to quantitatively predict and assess the consequences of these anatomical changes on tongue mobility and speech production after surgery.

## MATERIAL

The project is based on the use of a 3D biomechanical model of the tongue originally developed to study speech production in non pathological conditions [3,4].

This model represents the tongue as a Finite Element Structure with hexahedral elements. Based on indentation measurements on a fresh cadaver tongue, the hyperelastic properties account for the stress-strain relations of tongue tissues [5]. Muscles are represented within the finite element structure by specific subsets of elements, whose stress-strain relation varies with muscle activation. Muscle forces are applied to specific nodes of the structure via macrofibres that represent the main directions of muscle fibres in the different parts of the tongue. The tongue model is inserted in a vocal tract including mandible, palate and pharyngeal walls (figure 1, left panel).

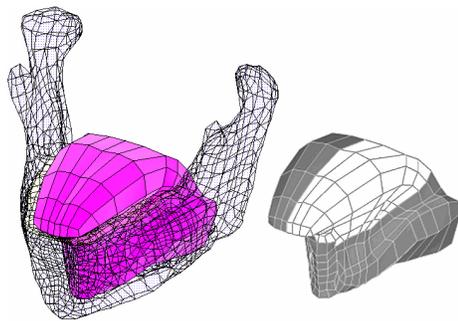

*Figure 1: Finite Element tongue model (left) with modifications in order to simulate hemiglossectomy and reconstruction with a flap (right, white elements of the mesh).*

Two examples of tongue surgery, which are quite common in the treatment of cancers of the oral cavity, can be modelled: hemiglossectomy and large resection of the mouth floor.

Most of the tongue cancers occur on its lateral border at the junction between the middle and posterior thirds. In these cases, for tumours with diameters as large as 2 cm the hemiglossectomy is the recommended surgical treatment. Cancer of the anterior floor of the mouth can also involve the ventral tongue or can extend along the lingual nerve, submandibular duct, or into the lingual cortex of the mandible. In that case, the resection sacrifices the muscular sling under the mucosa and can extend to the alveolar ridge, or interrupt the symphyseal mandible. At the back, the resection can include the ventral tongue. The simulation of an enlarged resection of the floor of the mouth has been described in Buchaillard *et al.* [6]. In that case, the anterior part of the genioglossus muscle was removed as well as the totality of two major muscles of the mouth floor, namely the geniohyoid and the mylohyoid muscles.

This paper focuses on the simulation of a left hemiglossectomy. The tongue model is divided lengthwise along the tongue septum, from the apex to the circumvallate papillae, and one half of the mobile part of the model is excised. The root and base of the tongue model are kept intact. Different kinds of flaps with various biomechanical properties can be used to reconstruct the tongue model (figure 1, right panel).

Three cases were studied. First we simply implemented a reconstruction with a flap having exactly the same biomechanical properties as the passive tissues of a healthy tongue. Then, flaps with a stiffness 5 times smaller and 6 times higher were simulated. These flaps are totally inactive during the tongue activation.

Figure 2 plots a frontal view of tongue deformations simulated once the styloglossus muscle is activated, i.e. when the tongue has to be pulled upwards and backwards in the direction of the velum. Such simulations show that some of the clinically observed consequences of tongue surgery, including partial resection and reconstruction with a flap, on tongue mobility, can be accounted for by the model. Indeed, whereas in normal conditions (figure 2a), the tongue movements are coarsely symmetric (pair styloglossus muscles are activated simultaneously), this symmetry is broken after a hemiglossectomy. Indeed, when the left-side of the tongue is removed and reconstructed (Figure 2(b-d)), an asymmetry of the tongue shaping is observed, characterized by a torsion of the tongue apex and of the tongue dorsum, and by an important deviation of the apex from the midsagittal plane.

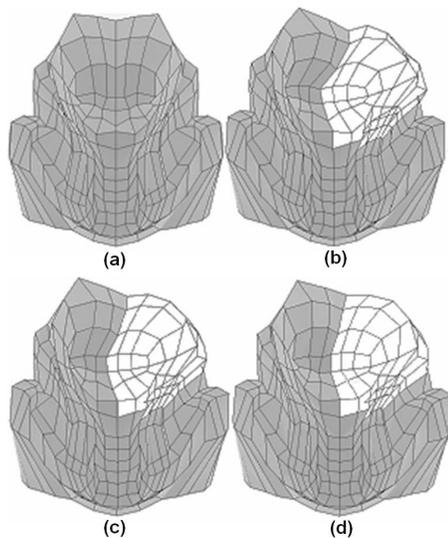

*Figure 2: Simulation of a left hemiglossectomy on the tongue model with activation of the styloglossus (frontal views of the normal model (a), and left-reconstructed models with a flap of increasing stiffness:* x0.2 *(b),* x1 *(c),* x6 *(d))*

It is also important to note that the physical properties of the flap have a strong impact on the amplitude of the tongue torsion and consequently on the maximal elevation of the tongue in the velar region. Indeed, the torsion decreases when the stiffness of the flap increases (figure 2 (b-d)), assuming that it seems more difficult to control the positioning of the surface of the tongue when the stiffness of the flap is small.

## CONCLUSION

In this paper, the ability of a 3D biomechanical model of the oral cavity to qualitatively predict the consequences of tongue surgery on tongue movements was studied, according to the size and location of the tissue loss and the nature of the flap used by the surgeon

The potential role of the mechanical characteristics of the flap on tongue mobility was shown and quantitatively evaluated. It appears that controlling flap stiffness could help preserving tongue mobility.

In conclusion, using such a model should represent a significant improvement in planning tongue surgery systems, in order to preserve as much as possible tongue mobility and therefore the patients' quality of life.

# *BIBLIOGRAPHY*

**[1] DELEYIANNIS FWB, WEYMULLER Jr. A, COLTRERA MD.**
*Quality of Life of Disease-Free Survivors of Advanced (Stage III or IV) Oropharyngeal Cancer.*
Head & Neck.1997; 19:466-73.

**[2] FURIA CLB, KOWALSKI LP, LATORRE MRDO, ANGELIS EC, MARTINS NMS, BARROS APB ET AL.**
*Speech Intelligibility After Glossectomy and Speech Rehabilitation.*
Archives of Otolaryngology and Head Neck Surgery. 2001; 127:877-883.

**[3] GERARD J-M, PERRIER P. & PAYAN Y.**
*3D biomechanical tongue modelling to study speech production*
In: J. Harrington & M. Tabain Editors, Speech Production: Models, Phonetic Processes, and Techniques. Psychology Press: New-York, USA; 2006. p. 85-102.

**[4] BUCHAILLARD S., PERRIER P. & PAYAN Y.**
*A 3D biomechanical vocal tract model to study speech production control: How to take into account the gravity?*
Proceedings of the 7th International Seminar on Speech Production, Ubatuba, Brazil; 2006. p. 403-10.

**[5] GERARD J-M, OHAYON J., LUBOZ V., PERRIER P. & PAYAN Y.**
*Non linear elastic properties of the lingual and facial tissues assessed by indentation technique. Application to the biomechanics of speech production*
Medical Engineering & Physics, 2005, Vol. 27 (10), pp. 884-892.

**[6] BUCHAILLARD S., BRIX M., PERRIER P. & PAYAN Y.**
*Simulations of the consequences of tongue surgery on tongue mobility: Implications for speech production in post-surgery conditions.*
International Journal of Medical Robotics and Computer Assisted Surgery, 2007 (in press).